\documentclass[prl,twocolumn,showpacs,nobibnotes,floatfix,superscriptaddress]{revtex4}

\usepackage{graphicx,eucal}

\begin{document}

\title{Matrix Element Randomness, Entanglement, and Quantum Chaos}

\author{Yaakov S. Weinstein}
\thanks{To whom correspondence should be addressed}
\email{weinstei@dave.nrl.navy.mil}
\author{C. Stephen Hellberg}
\email{hellberg@dave.nrl.navy.mil}
\affiliation{Center for Computational Materials Science, Naval Research Laboratory, Washington, DC 20375}

\begin{abstract}
We demonstrate the connection between an operator's matrix element distribution
and entangling power via numerical simulations of random, pseudo-random, 
and quantum chaotic operators. Creating operators with a random distribution 
of matrix elements is more difficult than creating operators that reproduce 
other statistical properties of random matrices. Thus, operators that 
fulfill many random matrix statistical properties may not generate states of 
high multi-partite entanglement. To quantify the randomness of various 
statistical distributions and, by extension, entangling power, we use 
properties of interpolating ensembles that transition between integrable 
and random matrix ensembles. 
\end{abstract}

\maketitle

Random matrices have long been used as statistical models for a host of complex
systems in areas of physics ranging from quantum dots to field 
theory \cite{RMT}. Originally introduced by Wigner to predict the energy 
level spectrum of heavy atomic nuclei \cite{Wigner}, random
matrices were later applied to characterize quantum chaotic systems 
\cite{BGS,Zyc,Haake}. More recently, random matrices have been found 
essential in various protocols of quantum computation and communication
including quantum state superdense coding \cite{Aram}, remote state 
preparation \cite{Bennet}, data hiding schemes \cite{Hayden}, and entangled 
state production \cite{Scott}.

Given the similarities between quantum chaotic operators and random 
matrices, and the aptitude of random matrices for producing entangled states, 
several authors have proposed entanglement production as an indicator of 
quantum chaos \cite{L1,FNP,MS}. While there has been some debate \cite{TFM}, 
substantial numerical evidence suggests that quantum systems with underlying 
chaotic classical dynamics produce entanglement at a faster rate than 
systems with underlying regular dynamics.  

In this work we explore the distribution of unitary operator matrix element
amplitudes, a statistical criterion of random matrices \cite{Zyc1}.
We show that the matrix elements are key to entanglement production. 
Furthermore, we demonstrate that, in creating an operator, it is more 
difficult to reproduce the statistical properties of the elements of random 
matrices than other random matrix properties. Evidence is provided from matrix 
ensembles that transition between integrable and random matrices, known 
as the interpolating ensembles, and pseudo-random matrices. For these 
classes of matrices nearest neighbor eigenvalue spacing and eigenvector 
element distributions converge to that of random matrices more quickly 
than the matrix element distribution. The entanglement production of 
these operators is correspondingly slower in converging to that of 
random matrix production. We also use interpolating ensembles properties
to quantify the randomness of a given statistical distribution and, 
by extension, entanglement production. Finally, we discuss the role 
of the matrix element distribution in the entanglement production of quantum 
chaotic operators: why the entanglement produced increases with time, and 
how the entanglement production differs from non-chaotic operators. 

The circular ensembles of random unitary matrices were introduced by Dyson
\cite{Dyson} as alternatives to the Gaussian ensembles of random Hermitian
matrices \cite{Wigner,Mehta}. As with the Gaussian ensembles, the circular
ensembles play a central role in modeling complex quantum systems.
Unlike the Gaussian ensembles, however, elements of the unitaries are not 
independent random variables \cite{Zyc2} and are thus more difficult to 
generate. Nevertheless, matrices of the circular unitary ensemble 
(CUE), the space of all unitary matrices, can be generated by taking the 
eigenvectors of a Hermitian matrix belonging to the Gaussian unitary 
ensemble (GUE), multiplying each by a random phase, and using the resulting 
vectors as the matrix columns \cite{Zyc1}. 

The squared modulus or amplitude of the CUE matrix elements follow a 
distribution equal to that of GUE eigenvector element amplitudes. Let 
$c^l_k$ denote the $k$th component of the $l$th GUE eigenvector. 
The distribution of amplitudes $\eta = |c^l_k|^2$ in the limit of infinite 
Hilbert space dimension, $N \rightarrow \infty$, after rescaling to a unit 
mean is $P_{GUE}(y) = e^{-y}$, where $y = N\eta$ \cite{Zyc}. Since $\eta$ is 
unchanged when multiplied by a phase, the distribution, $P_{CUE}(x)$, 
of the rescaled amplitude of CUE matrix elements $x$, is equal to $P_{GUE}(y)$.

CUE matrices can also be generated based on the Hurwitz parameterization 
using elementary unitary transformations, $E^{(i,j)}(\phi,\psi,\chi)$ 
with non-zero elements
\begin{eqnarray}
\label{E1}
E_{kk}^{(i,j)} &=& 1, \;\;\;\; k = 1, ... , N, \;\;\;\; k \neq i,j \nonumber\\
E_{ii}^{(i,j)} &=& e^{i\psi}\cos\phi, \;\;\;\;\;\;\;\;\; E_{ij}^{(i,j)} = e^{i\chi}\sin\phi \nonumber\\
E_{ji}^{(i,j)} &=& -e^{-i\chi}\sin\phi, \;\;\;\; E_{jj}^{(i,j)} = e^{-i\psi}\cos\phi
\end{eqnarray}
to construct $N-1$ composite rotations
\begin{eqnarray}
E_1 &=& E^{(N-1,N)}(\phi_{01},\psi_{01},\chi_1) \nonumber\\
E_2 &=& E^{(N-2,N-1)}(\phi_{12},\psi_{12},0)E^{(N-1,N)}(\phi_{02},\psi_{02},\chi_2) \nonumber\\
\dots\nonumber\\
E_{N-1} &=& E^{(1,2)}(\phi_{N-2,N-1},\psi_{N-2,N-1},0)\times\nonumber\\
 & & E^{(2,3)}(\phi_{N-3,N-1},\psi_{N-3,N-1},0)\times\nonumber\\
 &\dots& E^{(N-1,N)}(\phi_{0,N-1},\psi_{0,N-1},\chi_{N-1})
\end{eqnarray}
and, finally, $U_{CUE} = e^{i\alpha}E_1E_2\dots E_{N-1}$. Angles $\psi$, 
$\chi$, and $\alpha$ are drawn uniformly from the intervals
\begin{equation}
\label{E3}
0\leq \psi_{rs} \leq 2\pi, \;\;\;\;\;\; 0\leq \chi_{s} \leq 2\pi, \;\;\;\;\;\;
0\leq \alpha \leq 2\pi,
\end{equation}
and $\phi_{rs} = \sin^{-1}({\xi_{rs}}^{1/(2r+2)})$, with $\xi_{rs}$ drawn 
uniformly from 0 to 1 \cite{Zyc2}. Note that the $2\times2$ block 
$E^{(i,j)}_{m,n}$ with $m,n = i,j$ and $r = 0$ is a random SU(2) rotation 
with respect to the Haar measure.

An advantage of the above method is that it allows for a one-parameter 
interpolation between diagonal matrices with uniform, independently 
distributed elements and CUE \cite{Zyc3}. This is done by drawing the 
above parameters from constricted intervals
\begin{equation}
\label{delta}
0 \leq \psi_{rs} \leq 2\pi\delta, \;\;\;\;\;\; 0\leq \chi_s \leq 2\pi\delta, \;\;\;\;\;\; 0 \leq \alpha \leq 2\pi\delta, 
\end{equation}
and setting $\phi_{rs} = \sin^{-1}(\delta{\xi^j_i}^{1/(2r+2)})$ with 
$\xi_{rs}$ now drawn from 0 to $\delta$. The whole is 
multiplied by a diagonal matrix of random phases drawn uniformly from 0 to 
$2\pi$. The parameter $\delta$ ranges from 0 to 1 and provides a smooth 
transition between the diagonal circular Poisson ensemble (CPE) and CUE 
for nearest neighbor eigenvalue and eigenvector statistics \cite{Zyc3}. 

\begin{figure}
\includegraphics[height=5.8cm, width=8cm]{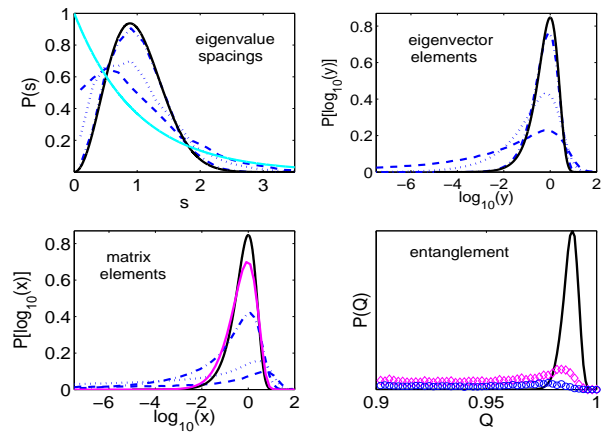}
\caption{\label{CUEdelta}
(Color online) Distributions of nearest neighbor eigenvalue spacings 
(upper-left) and eigenvector element amplitudes (upper-right) for matrices 
of the interpolating ensembles with $\delta = .1$ (dashed), .5 (dotted) 
and .9 (chained). The matrix element plot (lower-left) also includes the 
distribution for $\delta = .98$ (light solid line). For this $\delta$ the 
eigenvalue and eigenvector distributions are indistinguishable from 
random (solid line), but the matrix element distribution converges 
much more slowly. This is manifest in the 
entanglement generated by operating with 100 8-qubit matrices on all 
computational basis states, shown on the lower right for $\delta = .9$ (o), 
and $.98$ (diamonds).
}
\end{figure}

The interpolating ensembles matrix element amplitude distribution, however, 
does not transition smoothly with $\delta$. Rather, the matrix element 
randomness lags behind that of the other criteria such that even large 
$\delta$ induces matrix element distributions noticeably different from 
$P_{CUE}(x)$ as in figure \ref{CUEdelta}. 

A practical measure of multi-partite entanglement is the average bipartite 
entanglement between each qubit and the rest of the system \cite{Meyer,Bren2},
\begin{equation}
Q = 2-\frac{2}{n}\sum^n_{j=1}Tr[\rho_j^2],
\end{equation}
where $\rho_j$ is the reduced density matrix of qubit $j$.
An operators' lack of matrix element randomness causes the distribution of 
$Q$, for computational basis states evolved under one iteration of 
the operator, to diverge sharply from $P_{CUE}(Q)$, 
the distribution expected for states evolved by CUE matrices. 
For the interpolating ensembles a high value of $\delta$, such as 
$\delta = .98$, yields eigenvalue and eigenvector distributions extremely 
close to CUE. $P(Q)$, however, is still far from
$P_{CUE}(Q)$ due to lack of randomness in the matrix elements as seen in 
figure \ref{CUEdelta}.

The $\delta$ characterized distributions of the interpolating ensembles 
provide a quantifiable randomness measure for not fully random distributions.
Such a measure provides objectivity in reporting the distance a given 
distribution is from the random one and allows the randomness of 
different statistical criteria to be compared. Not all non-random 
statistics will follow one of these intermediate distributions, but we 
find them useful nevertheless. 

The randomness of an operator's matrix element distribution can be quantified 
by fitting it with an interpolating ensemble eigenvector distribution. Since 
an operators' matrix element distribution determines the amount of 
entanglement generated, reporting $\delta$ of the eigenvector distribution 
best fit provides a measure of entangling power. An operator or class of 
operators whose matrix element distribution conforms to a higher 
$\delta$-eigenvector distribution produces, on average, more entanglement. 
To generate a semi-random state with a given amount of entanglement, a 
quantum computer programmer can apply an operator with the desired matrix 
distribution $\delta$. We note that the interpolating ensemble matrix 
elements themselves are not well estimated by the $\delta$-eigenvalue 
distributions. This most likely stems from constrictions of the Euler angles, 
Eq. (\ref{delta}), which also leads to a non-smooth entanglement 
distribution.

To further explore the relationship between matrix element amplitude 
distributions, entanglement, and other statistical measures of randomness, 
we turn to pseudo-random operators \cite{RM,CRM,QCARM}. Despite being 
efficiently implementable on a quantum computer, pseudo-random operators 
have been shown to reproduce statistical properties of random operators. The 
pseudo-random operator algorithm is to apply $m$ iterations of the $n$ qubit 
gate: random SU(2) rotation to each qubit, then evolve the system via all 
nearest neighbor couplings \cite{RM}. A random SU(2) rotation with respect 
to the Haar measure is described by equations Eqs. (\ref{E1}) and (\ref{E3}).
The nearest neighbor coupling operator we use is
$U_{nnc} = e^{i(\pi/4)\sum^{n-1}_{j=1}\sigma_z^j\otimes\sigma_z^{j+1}}$,
where $\sigma_z^j$ is the $z$-direction Pauli spin operator.
The random rotations are different for each qubit and each iteration, but the
coupling constant is always $\pi/4$ to maximize entanglement. After the $m$
iterations, a final set of random rotations is applied.

\begin{figure}
\includegraphics[height=5.8cm, width=8cm]{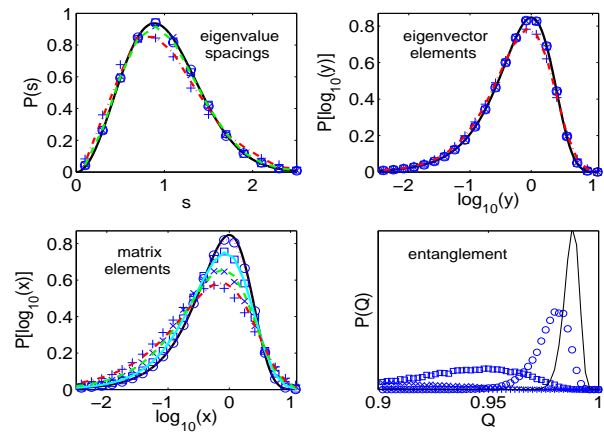}
\caption{\label{PR}
(Color online) Distribution of nearest neighbor eigenvalue spacings 
(top left), eigenvector elements (top right), matrix elements (bottom left), 
and $Q$ (bottom right) for 8-qubit pseudo-random maps of $m = 2$ (plus), 
4 (x), 8 (square), and 16 (o). The eigenvalue and eigenvector distributions 
converge to that of random matrices (solid lines) more quickly than the 
matrix element distribution. The eigenvalue spacing for $m = 2$ and $4$ 
are fitted by the $\delta = .8$ (chained line) and .94 (dashed line) 
eigenvalue distributions. The $m = 2$ eigenvector distribution is approximated 
by $\delta = .92$ (chained line). The distribution of matrix elements for 
the $m = 2, 4, 8$ and 16 pseudo-random operators can be approximated 
by $\delta = .7$ (chained line), .78 (dashed line), .88 (solid line), and 
.98 (dotted line) respectively. The low entanglement production is a direct 
outgrowth of the matrix element randomness lag. To approach $P_{CUE}(Q)$ 
requires $m \simeq 40$ \protect\cite{RM}.
}
\end{figure}

As $m$ increases, the statistical properties of the pseudo-random 
operators converge to those of random operators \cite{RM}. However, as 
seen in figure \ref{PR}, these properties do not converge at the same rate. 
For $n = 8, m = 2$ the eigenvalue spacing distribution is well 
approximated by the $\delta = .8$ interpolating ensemble eigenvalue spacing 
distribution, the eigenvectors by the $\delta = .92$ eigenvector element 
distribution, and the matrix elements by the $\delta = .7$ eigenvector element
distribution. Applying 100 $m = 2$ maps to all computational basis states
gives an average entanglement $\langle Q\rangle$ of .7004 compared to 
.9883 for CUE matrices. For $m = 4$ the eigenvalues are approximated 
by $\delta = .9$, while the eigenvectors are indistinguishable from 
the random distribution. The matrix elements are approximated
by a $\delta$ of .78 and $\langle Q\rangle = .8416$. Operators
of higher $m$ lead to eigenvalue and eigenvector distributions that are 
practically random. The matrix elements, however, lag behind with 
$\delta = .88$ and $\langle Q\rangle = .9339$ for $m = 8$ and $\delta = .98$
and $\langle Q\rangle = .9790$ for $m = 16$.

The above examples illustrate the difficulty in generating operators with a 
random distribution of matrix elements. In both cases increasing one parameter,
$\delta$ for the interpolating ensembles and $m$ for the pseudo-random 
operators, causes a convergence to CUE statistics. The 
convergence is relatively quick for eigenvalue and eigenvector distributions 
but much slower for the matrix element distribution. The correspondingly 
slow convergence of entanglement production to $P_{CUE}(Q)$ is evidence of 
the connection between the matrix element distribution and entanglement 
generation.

Quantum chaotic operators are known for their ability to produce
entanglement. Numerical simulations of two coupled subsystems demonstrate
the greater entanglement generation of chaotic versus regular quantum dynamics
\cite{L1,FNP,MS} and analytical results have been obtained through various 
methods \cite{TFM,L2,J}. Here, we are interested in a quantum chaotic 
operator's ability to produce entanglement on a quantum computer. Applying a 
chaotic operator once will not, in general, produce entanglement on par with 
random operators \cite{Scott,WGSH}. After a number of iterations, however, the 
average entanglement of initial computational basis states 
$\langle Q(t)\rangle$, where $t$ is the number of operator iterations, 
can approach that of random operators. We show 
how this is attributable to the matrix element distribution.

First, we note that an increase in $\langle Q(t)\rangle$ with time cannot be
due to eigenvector or eigenvalue statistics. The 
eigenvectors of an operator are unchanged as a function of $t$
and the eigenvalues become uncorrelated. The randomness of the matrix 
elements of chaotic operators, however, increases with $t$.
To demonstrate this assertion we revisit entanglement production 
of the quantum baker's map \cite{Scott} and explore other quantized 
chaotic maps. 

\begin{figure}
\includegraphics[height=5.8cm, width=8cm]{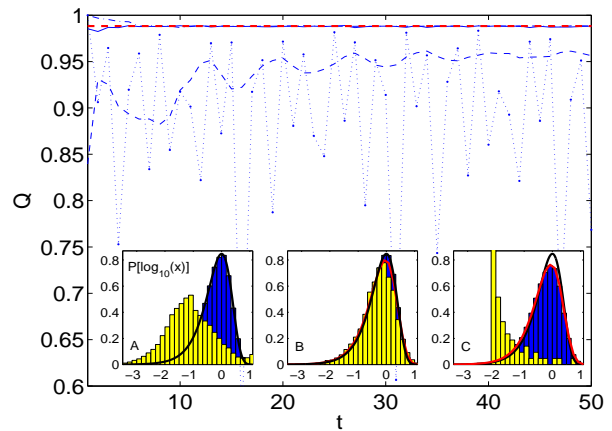}
\caption{\label{M2}
(Color online) Average entanglement, $Q$, over all 8-qubit initial 
computational basis states as a function of time for quantum sawtooth 
maps, $k = 1.5$ (chained line) and $k = -1.5$ (dotted line) and 
Harper maps, $\gamma = 1$ (solid line) and $\gamma = .1$ (dashed line), 
compared to the random matrix average (horizontal dashed line). The chaotic 
maps quickly approach the random matrix average while the regular maps do not. 
The insets show matrix element distributions for the regular (light) and 
chaotic (dark) sawtooth maps at $t = 50$ (A), the regular 
(light) and chaotic (dark) Harper maps at $t = 50$ (B), and $t = 1$ (light) 
and 100 (dark) of the baker's map (C). The matrix elements for the $t = 50$ 
regular Harper map is well approximated by the $\delta = .94$ eigenvalue 
distribution. 
}
\end{figure}

Initial computational basis states evolved under the 
quantum baker's map attain $Q$ values close to the random matrix average 
only at large $t$ \cite{Scott}. Figure \ref{M2}C shows that this
is due to the map's matrix element distribution. The baker's map matrix 
elements do not at all resemble $P_{CUE}(x)$, however, for $t = 100$ the 
distribution is well approximated by the $\delta = .9$ interpolating ensemble 
eigenvalue distribution. For the $n = 8$ baker's map $\langle Q(1)\rangle$ 
is only .3080, while $\langle Q(100)\rangle$ is .9597. More iterations
lead to increased matrix element randomness which causes greater entanglement 
generation.

Other examples of quantized chaotic maps are the quantum sawtooth 
map \cite{saw1,saw2},
$U_{saw} = \frac{e^{-i\pi/4}}{\sqrt{N}}e^{ik\pi m^2/N}e^{i\pi(n-m)^2/N}$,
and the quantum Harper map \cite{harper}, 
$U_H = e^{iN\gamma \cos(2\pi q/N)}e^{iN\gamma \cos(2\pi p/N)}$.
All elements of the chaotic, $k = 1.5$, and regular, $k = -1.5$, sawtooth 
maps have equal amplitude. Operating on any computational basis state 
yields a state of $Q = 1$. For the chaotic sawtooth the matrix element 
randomness increases with $t$, such that at $t = 50$ the matrix element 
distribution is practically $P_{CUE}(x)$ and $\langle Q(50)\rangle = .98826$. 
For the regular sawtooth the entanglement oscillates wildly 
as seen in figure \ref{M2}. This stems from the lack of an asymptotic 
randomness for the matrix elements.  

The matrix elements for the chaotic Harper, $\gamma = 1$, deviate only 
slightly from $P_{CUE}(x)$, and, thus, $\langle Q(1)\rangle =.9814$. For 
$t = 50$  there is the expected increase in matrix element randomness and 
$\langle Q(50)\rangle = .9882$. The regular Harper map, $\gamma = .1$, 
matrix element distribution and average entanglement also approach asymptotic 
limits as $t$ increases. However, these limits fall short of the random matrix
statistics. The matrix element distribution is well fit by the $\delta = .9$ 
interpolating ensemble eigenvector element distribution and the average 
entanglement $\langle Q(t\rightarrow\infty)\rangle \simeq .95$. 
We note that the matrix elements of the 100 times iterated baker's map and 
the $m = 8$ pseudo-random operators also follow the $\delta = .9$ 
distribution. Thus, $\langle Q(50)\rangle$ of the regular harper is very 
close to $\langle Q\rangle$ of these operators.

In conclusion, we have explored connections between a system's 
matrix element distribution and entanglement production. The interpolating 
ensembles demonstrate the difficulty in creating operators with randomly 
distributed matrix elements and their statistical properties provide a 
useful measure of randomness. This measure is used to compare and contrast the 
randomness of statistical distributions from pseudo-random and chaotic 
operators. The connection to matrix element randomness provides a random 
matrix basis for increased entanglement production of chaotic systems 
as a function of time and the greater entanglement production for chaotic 
versus non-chaotic systems. Though this need not always be the case, operators 
with similar matrix element distributions appear, on average, to 
produce states with a similar amount of entanglement.

The authors thank K. Zyczkowski for clarifying interpolating ensemble
generation. The authors acknowledge support from the DARPA QuIST 
(MIPR 02 N699-00) program. YSW acknowledges the support
of the National Research Council Research Associateship Program through
the Naval Research Laboratory. Computations were performed at the ASC
DoD Major Shared Resource Center.

\end{document}